# ChronoNet: A Deep Recurrent Neural Network for Abnormal EEG Identification

**Subhrajit Roy, Isabell Kiral-Kornek, and Stefan Harrer**
IBM Research – Australia, 60 City Rd, Southbank, VIC, 3006, Australia
{subhrajit.roy, isabeki, sharrer}@au1.ibm.com

## Abstract

Brain-related disorders such as epilepsy can be diagnosed by analyzing electroencephalograms (EEG). However, manual analysis of EEG data requires highly trained clinicians, and is a procedure that is known to have relatively low inter-rater agreement (IRA). Moreover, the volume of the data and the rate at which new data becomes available make manual interpretation a time-consuming, resource-hungry, and expensive process. In contrast, automated analysis of EEG data offers the potential to improve the quality of patient care by shortening the time to diagnosis and reducing manual error. In this paper, we focus on one of the first steps in interpreting an EEG session - identifying whether the brain activity is abnormal or normal. To address this specific task, we propose a novel recurrent neural network (RNN) architecture termed ChronoNet which is inspired by recent developments from the field of image classification and designed to work efficiently with EEG data. ChronoNet is formed by stacking multiple 1D convolution layers followed by deep gated recurrent unit (GRU) layers where each 1D convolution layer uses multiple filters of exponentially varying lengths and the stacked GRU layers are densely connected in a feed-forward manner. We used the recently released TUH Abnormal EEG Corpus dataset for evaluating the performance of ChronoNet. Unlike previous studies using this dataset, ChronoNet directly takes time-series EEG as input and learns meaningful representations of brain activity patterns. ChronoNet outperforms previously reported results on this dataset thereby setting a new benchmark. Furthermore, we demonstrate the domain-independent nature of ChronoNet by successfully applying it to classify speech commands.

## 1 Introduction

Electroencephalography (EEG) is a noninvasive method to measure brain activity through the recording of electrical activity across a patient's skull and scalp and is frequently used for the diagnosis and management of various neurological conditions such as epilepsy, somnipathy, coma, encephalopathies, and others. Despite having lower spatial resolution than brain imaging techniques such as magnetic resonance imaging (MRI) and computed tomography (CT), EEG is a popular diagnostics tool among physicians due to its excellent temporal resolution, low cost, and noninvasive nature [21].

As symptoms are not guaranteed to be present in the EEG signal at all times, the diagnosis of a neurological condition via EEG interpretation typically involves long-term monitoring or the recording of multiple short sessions. In this process, large amounts of data are generated that subsequently need to be manually interpreted by expert investigators. The relatively low availability of certified expert investigators and high volume of data make EEG interpretation a time-consuming process that can introduce a delay of hours to weeks in the patient's course of treatment. Introducing a certain level of automation to the EEG interpretation task could serve as an aid to neurologists by accelerating the reading process and thereby reducing workload. It is these reasons why automatic interpretation of EEG by machine learning techniques has gained popularity in recent times [7, 20].

When interpreting an EEG recording, first, an assessment is made as to whether the recorded signal appears to show abnormal or normal brain activity patterns as per [15]. This decision can influence which medication is being prescribed or whether further investigation is necessary. Typically, both, patterns in the recording and the patient's state of consciousness are being considered when deciding whether a recording shows a abnormal or normal EEG. Highly trained clinicians typically follow a complicated decision chart to make this distinction [15].

The motivation behind our work is to automate this first step of interpretation. We do so using a recently released dataset known as the TUH Abnormal EEG Corpus, which is the largest of its type to date [17] and freely available at [1]. Inspired by successess in time-domain signal classification, we explore recurrent neural network (RNN) architectures using the raw EEG time-series signal as input. This sets us apart from previous publications [15, 16, 19], in which the authors used both traditional machine learning algorithms such as k-nearest neighbour, random forests, and hidden markov models and modern deep learning techniques such as convolutional neural networks (CNN), however, did not use RNNs for this task.

Compared to the original studies using hand-engineered features [15, 16], we show that the combination of raw time series and RNNs eliminates the need to extract hand-crafted features and allows the classifier to automatically learn relevant patterns, surpassing their results by 3.51%. Taking inspiration from 1D convolution layers [8], gated recurrent units [4], inception modules [22], and densely connected networks [13], we build a novel deep gated RNN named ChronoNet which further increases accuracy by an additional 4.26%, resulting in an overall 7.77% improvement over results reported in [15, 16]. Moreover, compared to a recently published study showing state-of-the-art performance on this dataset, ChronoNet achieves 1.17% better results thereby setting a new benchmark for the TUH Abnormal EEG Corpus. In summary, the primary contributions of this paper are:

- Network Architecture: We use inception layers with exponentially varying kernel lengths for 1D convolution layers in combination with densely connected recurrent layers.
- Application: We achieve state-of-the-art performance in an EEG classification task and systematically showed how each component affects the performance.

By applying ChronoNet to the Speech Commands Dataset [25], we show its utility for general time series analysis beyond EEG interpretation.

## 2 Background and Theory

Raw EEG signals are temporal recordings that may exhibit patterns and periodicities at various time scales. A method that has successfully been used to classify time signals, for example speech, is the use of recurrent neural networks (RNNs). In this section, we first describe the working principle of RNNs. This will be followed by the description of a modern and sophisticated recurrent unit named gated recurrent unit (GRU) that is well suited to learning longer-term dependencies and correlations. We will then discuss the concept of inception modules and densely connected neural networks (concepts used in convolutional neural networks) which we will use for EEG data analysis to account for patterns emerging at different scales and for mitigation of vanishing gradients, respectively. This collection of principles and modules provides the essential basis for understanding the ChronoNet architecture proposed in Section 3.

### 2.1 Recurrent Neural Networks

RNNs are a family of neural networks for processing variable-length sequential data. A RNN maintains a recurrent hidden state whose activation at each time is dependent on that of the previous time step. More formally, given a sequence $\mathbf{x} = (x_1, x_2, ..., x_T)$, at each time step $t$, a RNN updates its recurrent hidden state $h_t$ based on the current input vector $x_t$ and the previous hidden state $h_{t-1}$ as follows:

$$h_t = \begin{cases} 0 & \text{if } t = 0 \\ g(h_{t-1}, x_t) & \text{otherwise,} \end{cases} \quad (1)$$

where $g$ is a nonlinear function.

In a classical RNN, the recurrent hidden unit of Equation( 1) is updated in the following way:

$$h_t = f(Wx_t + Uh_{t-1} + b), \quad (2)$$

where $f$ is a pointwise nonlinear activation function such as a logistic sigmoid function or a hyperbolic tangent function.

While Equation (2) allows an RNN to process sequences of arbitrary length, it has been observed that gradients of $f$ can grow or decay exponentially over long sequences during training [3]. This phenomenon makes it difficult for an RNN to learn long-term dependencies and correlations.

One way of tackling this issue is to design more sophisticated recurrent units which compute an affine transformation followed by a simple element-wise nonlinearity by using gating units. Two popular models in use are the long short-term memory (LSTM) [11, 9] and the gated recurrent unit (GRU) [4]. While it has been shown that LSTMs and GRUs significantly outperform classical RNNs, which one among the two performs better is yet to be conclusively shown [6]. In this paper, we use GRUs since they use fewer parameters than LSTMs and hence offer a faster training time while requiring data to generalize.

## 2.2 Gated Recurrent Unit

A gated recurrent unit (GRU) produces the current value of hidden state $h_t$ by performing a linear interpolation between an intermediate candidate hidden state $\tilde{h}_t$ derived from Equation (2) and the previous value of hidden state $h_{t-1}$. A GRU employs two gates: an *update gate* $z_t$ controlling the extent to which the previous state will be overwritten and a *reset gate* $r_t$ deciding how much of the previous state should be forgotten while computing the candidate hidden state. More formally, the GRU model can be presented in the following mathematical form:

$$h_t = (1 - z_t) \odot h_{t-1} + z_t \odot \tilde{h}_t \qquad (3)$$
$$\tilde{h}_t = g(W_h x_t + U_h(r_t \odot h_{t-1}) + b_h) \qquad (4)$$
$$z_t = \sigma(W_z x_t + U_z h_{t-1} + b_z) \qquad (5)$$
$$r_t = \sigma(W_r x_t + U_r h_{t-1} + b_r), \qquad (6)$$

where $g$ and $\sigma$ are nonlinear activation functions and $\odot$ denotes element-wise multiplication.

## 2.3 Inception Module

The inception module was proposed by Szegedy et al. [22] as a building block for the GoogLeNet architecture. Unlike traditional convolutional neural networks, the inception module uses filters of varied size in a convolution layer to capture features of different levels of abstraction. Processing visual information at different scales and aggregating them allows the network to efficiently extract relevant features. Typically, the module uses three filters of sizes $1 \times 1$, $3 \times 3$, and $5 \times 5$. Moreover, an alternative parallel path is also included which implements a $3 \times 3$ max-pooling operation.

However, naively introducing more filters in convolutional layers increases the number of parameters. This makes training of a network computationally more intensive compared to traditional CNNs. Hence, $1 \times 1$ filters are included in the inception module to implement "bottleneck layers" for dimensionality reduction.

## 2.4 DenseNet

DenseNet is a deep convolutional neural network architecture recently proposed by [13]. The main idea of DenseNet is that it connects each layer with every other layer in a feed-forward fashion. Each layer uses the feature maps of all its preceding layers as input and passes its own feature maps as input to all subsequent layers. Hence, while a traditional CNN with $L$ layers has $L$ connections, in DenseNet there are $L(L+1)/2$ direct connections.

DenseNet mitigates the problem of vanishing/exploding gradients that is observed in very deep networks [10]. It achieves that by providing short-cut paths for the gradients to pass during backpropagation. This allows the learning algorithm to choose the appropriate model complexity during training.

# 3 Methods

Inception modules (see Section 2.3) were originally proposed to enable a convolutional neural network to account for different abstraction layers in the context of image processing. Similarly, densely connected networks (see Section 2.4) were developed to address vanishing gradients due to backpropagation in deep convolutional neural networks.

As described previously, EEG data contains information across different scales in the time-domain. Furthermore, using deep RNN architectures might lead to the problem of vanishing or exploding gradients. If designed properly, advantages of inception modules and densely connected layers may, thus, equally apply to problems in the time-domain. In the following section, we will use the concepts of inception modules and densely connected networks to build novel recurrent neural networks architectures for time-series classification.

## 3.1 Convolutional Gated Recurrent Neural Network (C-RNN)

Considering the input is a time series, an obvious first approach is to stack multiple GRU layers as shown in Figure 1a. This popular architecture for handling sequential input data has led to state-of-the-art accuracy in various pattern recognition tasks, especially in natural language processing [24, 26].

However, when applied to relatively long input time-series data (as opposed to embedding vectors [18] in the case of natural language processing), this approach turns out to be computationally very intensive and time consuming to train. To solve this problem, data can be downsampled to an acceptable length before it is given as input to RNNs. However, using fixed values means that networks will not be able to adapt to the data at hand. In order to mitigate these problems, we used multiple 1D convolution (Conv1D) layers with strides larger than 1, enabling the network to learn to appropriately reduce the input signal automatically.

The resulting architecture (C-RNN) is a combination of Conv1D layers followed by stacked GRU layers [23, 5]. Conv1D layers have two advantages. First, they learn to sub-sample the signal and, thus, reduce the input vector's length as we move towards higher layers. This becomes particularly relevant when reaching GRU layers, which during training constitute the most computationally expensive part of the network. Second, Conv1D layers extract local information from neighbouring time points, a first step towards learning temporal dependencies. Following Conv1D layers, the GRU layers are responsible for capturing both short- and long-term dependencies.

The specific network used in this paper is presented in Figure 1b. The formats used throughout the paper to describe Conv1D and GRU layers are (layer name, filter length, number of filters, stride size) and (layer name, number of filters) respectively.

## 3.2 Inception Convolutional Gated Recurrent Neural Network (IC-RNN)

In the previous C-RNN architecture, each Conv1D layer had the capability to extract local information at only one time scale determined by a single fixed filter size, limiting the flexibility of the model. Since the rate of change of information in a time series depends on the task at hand, the filter size for each Conv1D layer would have to be hand-picked to fit the particular data.

To address this problem, taking inspiration from [22], we designed an architecture which expands upon C-RNN by including multiple filters of varying sizes in each Conv1D layer. This allows for the network to extract information over multiple time-scales. However, unlike [22], in IC-RNN, filter lengths used in the Conv1D layers were drawn from a logarithmic instead of a linear scale, leading to exponentially varying filter lenghts. Our experiments demonstrated that for the dataset considered in this paper, exponentially varying filter lengths lead to better performance. We speculate that this is because compared to images where relevant features vary in the same order of magnitude, in time series the range of timescales in which features exist is much wider. Note that, to the best of our knowledge, inception modules with exponentially varying filter sizes are reported for the first time in this paper. The specific configuration used in our experiments is shown in Figure 1c. A Filter Concat layer concatenates the incoming features along the depth axis.

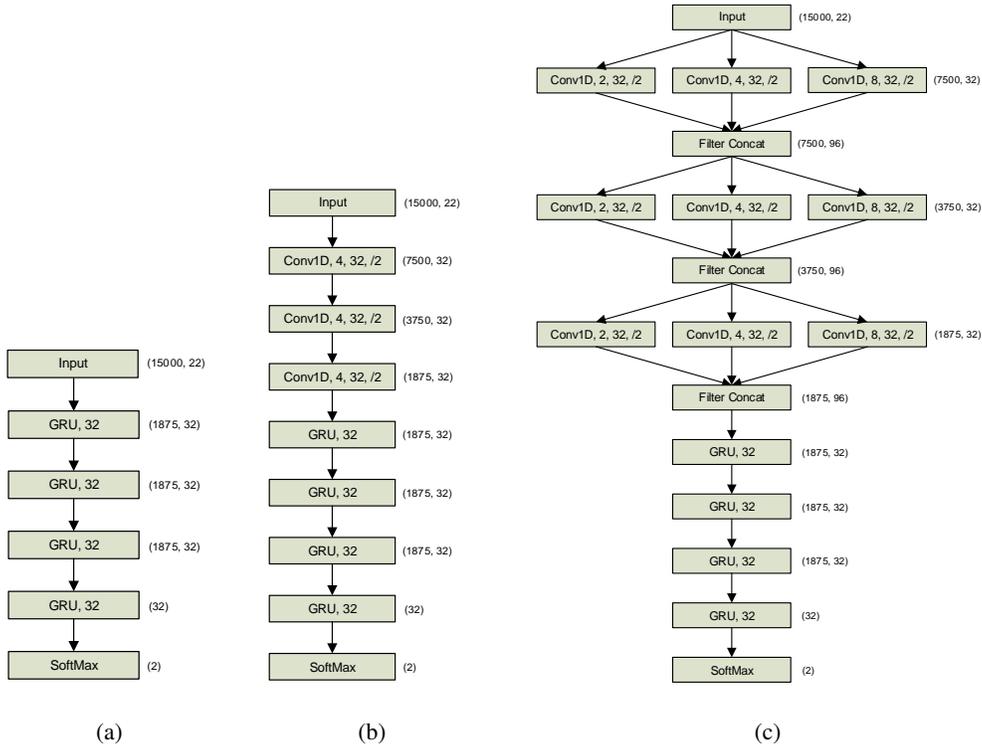

Figure 1: (a) A deep gated recurrent neural network which feeds the input directly into stacked GRU layers. (b) The convolutional gated recurrent neural network (C-RNN) which stacks multiple 1D convolution layers followed by a small number of GRU layers. (c) The Inception Convolutional Gated Recurrent Neural Network (IC-RNN) modifies the C-RNN (see Figure 1b) architecture by including multiple filters of exponentially varying lengths in the 1D convolution layers.

### 3.3 Convolutional Densely Connected Gated Recurrent Neural Network (C-DRNN)

The C-RNN architecture is not immune to the problem of *degradation* which sometimes impedes the training of very deep neural networks [10]. For simpler problems that do not need the full potential of the model complexity offered by a C-RNN, the optimization procedure may lead to higher training errors.

To tackle this issue, inspired by the DenseNet architecture proposed by [13] for CNNs, we incorporate skip connections in the stacked GRU layers of C-RNN to form the C-DRNN architecture. Each GRU layer is connected to every other GRU layer in a feed-forward fashion. Intuitively, skip connections will lead to GRU layers being ignored when the data demands a lower model complexity than offered by the entire network. The details of the network are shown in Figure 2a.

### 3.4 ChronoNet: Inception Convolutional Densely Connected Gated Recurrent Neural Network

Finally, we combine both modifications introduced for the previous two networks (IC-RNN and C-DRNN) with C-RNN to form the ChronoNet architecture. To the best of our knowledge, this is the first time this architecture has been reported. To summarize, ChronoNet is created by stacking multiple Conv1D layers followed by multiple GRU layers where each Conv1D layer has multiple filters of varying sizes and the stacked GRU layers are densely connected in a feed-forward manner.

The presence of multiple filters in the Conv1D layers allows ChronoNet to extract and combine features from different time scales. The optimum filter size for a Conv1D layer usually depends on both the task at hand and its relative position in the network. ChronoNet has the flexibility to explore multiple filter lengths for each Conv1D layer. On the other hand, densely connected GRU layers allow ChronoNet to mitigate the problem of *degradation* of training accuracy caused by vanishing or exploding gradients. This potentially enables the creation of very deep variants of ChronoNet for more complex tasks. Moreover, dense connections also strengthen feature propagation and encourage feature reuse in the GRU layers. The network we designed for the abnormal EEG classification task considered in this paper is depicted in Figure 2b.

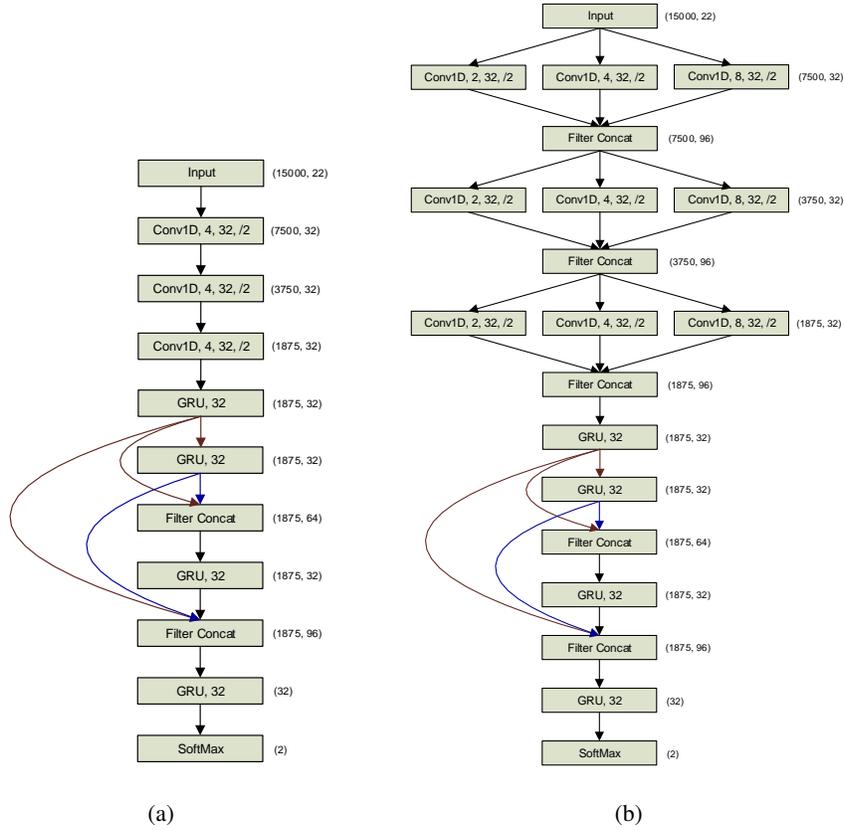

Figure 2: (a) The Convolutional Densely Connected Gated Recurrent Neural Network (C-DRNN) architecture includes dense connections in the recurrent stage i.e the output of each GRU layer is given as input to every other GRU layers in a feed-forward manner. (b) Proposed ChronoNet architecture which includes both multiple filters of exponentially varying lengths in the 1D convolution layers and dense connections within the GRU layers.

## 4 Experiments

In this section, we briefly describe the dataset, the data augmentation technique we used, and present and discuss the obtained results.

### 4.1 Data Selection

In this paper, we primarily focused on the TUH Abnormal EEG Corpus [16], which contains EEG records that are annotated as either clinically abnormal or normal. The TUH Abnormal EEG Corpus is a subset of the TUH EEG Corpus [17] which is the world's largest publicly available database of clinical EEG data. The TUH EEG Corpus comprises 23257 EEG sessions recorded over 13551 patients. In the entire dataset, almost 75% of the data represent abnormal EEG sessions. The TUH EEG Abnormal Corpus was formed by selecting a demographically balanced subset of the TUH EEG Corpus through manual review that consisted of 1488 abnormal and 1529 normal EEG sessions, respectively. These sets were further partitioned into a training set (1361 abnormal/1379 normal), and a test set (127 abnormal/150 normal).

| –        | Abnormal | Normal |
|----------|----------|--------|
| Training | 14971    | 15169  |
| Test     | 127      | 150    |

Table 1: Size of training and test set used in experiments of Section 4.3.

### 4.2 Data Preparation

TUH Abnormal EEG Corpus consists of EEG sessions recorded according to the 10/20 electrode configuration [12]. We converted the recorded EEG signal into a set of montages or differentials based on guidelines proposed by the American Clinical Neurophysiology Society [2]. In this paper, we used the transverse central parietal (TCP) montage system for accentuating spike activity [16]. Note that we did not extract any hand-engineered features from the dataset because we envisioned that the deep RNNs used in this paper will be able to automatically extract relevant features and learn meaningful representations.

In the original study [16], the authors noted that neurologists can accurately classify an EEG session into either abnormal or normal by only examining the initial portion of the signal. This motivated the authors to build machine learning algorithms that can classify an EEG signal by taking only the first minute of data as input. Hence, training and test set were generated by extracting the first minute from the available EEG sessions. Note that during testing only the first minute was used to enable a fair comparison of the classifier to human-level performance. Using only the first minute to create the training set, on the other hand, was a design choice motivated by the fact that the first minute might be most representative of the test set. Once electrodes are placed on the scalp and data recording starts, impedances and therewith the signal will gradually change due to external factors such as slowly drying conductive paste. To have a fair comparison with [16], we trained our model only on the first minute and report the obtained results in 4.3.

However, using the above method significantly limits the amount of data that can be used for training. This results in two problems. First, deep learning is a data hungry technique and performance substantially increases as more data is included in the training set. Second, when applied to small datasets, RNNs have a tendency to quickly overfit, an effect that intensifies as networks become deeper, as is the case for those considered in this paper. To not limit ourselves unnecessarily, we analyzed the effect of including more than just the first minute from the training sessions. This was done by choosing a random subset of sessions from the original training by exclusion of any samples later used for testing. The resulting sets were then further divided into smaller training and test sets. Separate models were trained for each minute of these training sets. We analysed the performance of these models on the first minute of the intermediate small test sets.

The outcome of this experiment demonstrated that we can use up to 11 minutes of data from the training EEG sessions without performance degradation. This led to a 11-fold increase in our training data as compared to the method used in [16]. The sizes of the final training and test set used in our experiments are shown in Table 1.

In this dataset, most recordings were done with a sampling frequency of 250 Hz. Where this was not the case, sessions were resampled to 250 Hz. An input vector to the network was 1 minute long, thus, consisting of 15000 time points.

### 4.3 Results

We used the dataset described above to train the four deep recurrent neural network architectures presented in Section 3. Networks were trained using the adaptive moment estimation optimization [14] algorithm with a learning rate of 0.001. Moreover, we used a batch size of 64 and trained the networks for 500 epochs. Table 2 lists mean accuracies of 5 repetitions of these experiments. Results reported to date on this dataset are included for comparison. In [16], the author explored various machine- and deep learning algorithms and observed that best performance is obtained when frequency features extracted from the input time-series signal are fed into a convolutional neural network [16] (CNN-MLP in Table 2). Furthermore, in [19] the authors used a deep convolutional neural network built by automatic hyperparameter search (DeepCNN in Table 2)) and reported the best accuracy to date.

| – | Training accuracy | Testing accuracy |
|---|---|---|
| C-RNN | 83.58% | 82.31% |
| IC-RNN | 86.93% | 84.11% |
| C-DRNN | 87.20% | 83.89% |
| ChronoNet | **90.60%** | **86.57%** |
| CNN-MLP | N/A | 78.80% |
| DeepCNN | N/A | 85.40% |

Table 2: Performance comparison of the four deep recurrent neural networks described in Section 3 and results reported in [16] (see CNN-MLP) and [19] (see DeepCNN).

Table 2 clearly depicts that the deep recurrent neural architectures explored in this paper outperform the results shown in the original study [16] using CNN-MLP. It is important to note that in contrast to CNN-MLP, the proposed architectures do not rely on hand-crafted features. Moreover, we see that C-RNN, IC-RNN, C-DRNN, and ChronoNet are surpassing best accuracy reported in [16] by 3.51%, 5.31%, 5.09%, and 7.77%, respectively. Furthermore, compared to the recently published state-of-the-art performance [19], ChronoNet shows 1.17% better accuracy. Out of the four recurrent architectures, ChronoNet achieves both the best training and testing accuracy. This shows that the combined positive effect of including multiple filters in Conv1D layers and incorporating dense connections in the GRU layers is more pronounced than using either one or none of them. Moreover, our experiments showed that ChronoNet yields similar performance (86.64%, averaged over 5 runs) when GRUs are replaced by LSTMs, however, networks with LSTM units took longer time to train than their GRU counterpart.

The training and test dataset was pre-split in the TUH Abnormal EEG corpus in a way such that each set is demographically balanced (gender and age) and no patient appears in both the training and testing set. To demonstrate that the network is not overfitting the hyper parameters on the test set, we combine the training and test set provided in the TUH Abnormal EEG Corpus and perform a 5-fold cross-validation to provide test accuracy for the proposed architecture. We achieve a 86.14% accuracy with the 5-fold cross-validation approach.

Note that the number of EEG records used in the training set is the same as the number used in other works on this dataset. While the original study [16] used only the first minute, we discovered that more than the first minute can be included in the training set. If trained on just the first minute i.e. when the training set is exactly same as used in other work, ChronoNet achieves an accuracy of 85.27% (averaged over 5 runs) which is 6.47% better than [16].

To demonstrate that exponentially varying filter sizes in the Conv1D layers of ChronoNet are a necessary component, two experiments were performed. First, shorter (compared to the longest 1D convolution filter used in ChronoNet) linearly varying filters of lengths 3, 5, and 7 were implemented. As a result, training and testing accuracies fall to 89.15% and 85.12%, respectively. Second, longer but linearly varying filters of lengths of 14, 16, and 18 were implemented. While training accuracy increased to 91.25%, the testing accuracy was reduced to 85.92%. We speculate that in both cases features extracted by the network are not sufficiently diverse, and furthermore, in the later case, the increased model complexity leads to overfitting.

### 4.4 Beyond EEG Identification

ChronoNet is designed to efficiently find patterns on different time scales in temporal data with its main advantages lying in its flexibility and adaptability. While in this paper we primarily concentrated on applying ChronoNet for classifying abnormal/normal EEG, it can also be applied to the broader domain of time-series classification. In a preliminary study, we are using ChronoNet to solve a speech classification task, using a recently released dataset known as the Speech Commands Dataset [25]. The dataset consists of one-second long utterances of 30 short words, spoken by thousands of different people. The sizes of the training, validation, and test set are 64721, 6798, and 6835 samples respectively. Using the exact same architecture as shown in Figure 2b, we achieved a testing accuracy of 92.84% (averaged over 5 runs) for this 30 class problem. We intend to undertake a larger study on different time-domain datasets in the near future.

## 5 Conclusion

Determining whether an EEG recording shows abnormal or normal brain activity is often the first step in the diagnosis of a neurological condition. Since manual interpretation of EEG is an expensive and time-consuming process, any classifier that automates this first distinction will have the potential to reduce delays in treatment and to relieve clinical care givers. We introduce ChronoNet, a novel network architechture that is designed to be flexible and adaptable and, thus, uniquely suited for the analysis of EEG time-series data. This novel RNN architecture outperforms the best previously reported accuracy on the dataset used by 1.17%, setting a new benchmark. To demonstrate its general applicability to time-series data, we further demonstrate that ChronoNet can successfully classify speech.

## References

[1] Temple university EEG corpus. *Link: https://www.isip.piconepress.com/projects/tuh_eeg/*.

[2] Jayant N. Acharya, Abeer J. Hani, Partha D. Thirumala, and Tammy N. Tsuchida. American Clinical Neurophysiology Society Guideline 3: A Proposal for Standard Montages to Be Used in Clinical EEG. *J Clin Neurophysiol*, 33(4):312–316, August 2016.

[3] Y. Bengio, P. Simard, and P. Frasconi. Learning Long-term Dependencies with Gradient Descent is Difficult. *Trans. Neur. Netw.*, 5(2):157–166, March 1994.

[4] Kyunghyun Cho, Bart van Merrienboer, Dzmitry Bahdanau, and Yoshua Bengio. On the Properties of Neural Machine Translation: Encoder-Decoder Approaches. *arXiv:1409.1259v2*, September 2014.

[5] Keunwoo Choi, George Fazekas, Mark B. Sandler, and Kyunghyun Cho. Convolutional recurrent neural networks for music classification. *CoRR*, abs/1609.04243, 2016.

[6] Junyoung Chung, Caglar Gulcehre, KyungHyun Cho, and Yoshua Bengio. Empirical Evaluation of Gated Recurrent Neural Networks on Sequence Modeling. *arXiv:1412.3555 [cs]*, December 2014. arXiv: 1412.3555.

[7] Meysam Golmohammadi, Saeedeh Ziyabari, Vinit Shah, Silvia Lopez de Diego, Iyad Obeid, and Joseph Picone. Deep Architectures for Automated Seizure Detection in Scalp EEGs. *arXiv:1712.09776 [cs, eess, q-bio, stat]*, December 2017. arXiv: 1712.09776.

[8] Ian Goodfellow, Yoshua Bengio, and Aaron Courville. Deep Learning | The MIT Press, 2016.

[9] Alex Graves. Generating Sequences With Recurrent Neural Networks. *arXiv:1308.0850 [cs]*, August 2013. arXiv: 1308.0850.

[10] Kaiming He, Xiangyu Zhang, Shaoqing Ren, and Jian Sun. Deep Residual Learning for Image Recognition. *arXiv:1512.03385 [cs]*, December 2015. arXiv: 1512.03385.

[11] Sepp Hochreiter and Jürgen Schmidhuber. Long Short-Term Memory. *Neural Computation*, 9(8):1735–1780, November 1997.

[12] Richard W. Homan. The 10-20 electrode system and cerebral location. *American Journal of EEG Technology*, 28(4):269–279, 1988.

[13] Gao Huang, Zhuang Liu, Kilian Q. Weinberger, and Laurens van der Maaten. Densely Connected Convolutional Networks. *arXiv:1608.06993*, August 2016.

[14] Diederik P. Kingma and Jimmy Ba. Adam: A Method for Stochastic Optimization. *arXiv:1412.6980 [cs]*, December 2014. arXiv: 1412.6980.

[15] S. López, G. Suarez, D. Jungreis, I. Obeid, and J. Picone. Automated Identification of Abnormal Adult EEGs. *IEEE Signal Process Med Biol Symp*, 2015, December 2015.

[16] Silvia López. Automated Interpretation of Abnormal Adult Electroencephalograms. *MS Thesis, Temple University. Link: https://www.isip.piconepress.com/publications/ms_theses/2017/abnormal/*, 2017.

[17] Iyad Obeid and Joseph Picone. The Temple University Hospital EEG Data Corpus. *Front Neurosci*, 10, May 2016.

[18] Jeffrey Pennington, Richard Socher, and Christopher D. Manning. GloVe: Global Vectors for Word Representation. In *Empirical Methods in Natural Language Processing (EMNLP)*, pages 1532–1543, 2014.

[19] Robin Tibor Schirrmeister, Lukas Gemein, Katharina Eggensperger, Frank Hutter, and Tonio Ball. Deep learning with convolutional neural networks for decoding and visualization of EEG pathology. *CoRR*, abs/1708.08012, 2017.

[20] Robin Tibor Schirrmeister, Jost Tobias Springenberg, Lukas Dominique Josef Fiederer, Martin Glasstetter, Katharina Eggensperger, Michael Tangermann, Frank Hutter, Wolfram Burgard, and Tonio Ball. Deep learning with convolutional neural networks for EEG decoding and visualization. *Human Brain Mapping*, 38(11):5391–5420, November 2017. arXiv: 1703.05051.

[21] S. J. M. Smith. EEG in the diagnosis, classification, and management of patients with epilepsy. *Journal of Neurology, Neurosurgery & Psychiatry*, 76(suppl 2):ii2–ii7, 2005.

[22] C. Szegedy, Wei Liu, Yangqing Jia, P. Sermanet, S. Reed, D. Anguelov, D. Erhan, V. Vanhoucke, and A. Rabinovich. Going deeper with convolutions. In *2015 IEEE Conference on Computer Vision and Pattern Recognition (CVPR)*, pages 1–9, June 2015.

[23] Duyu Tang, Bing Qin, and Ting Liu. Document modeling with gated recurrent neural network for sentiment classification. In *Proceedings of the 2015 Conference on Empirical Methods in Natural Language Processing*, pages 1422–1432. Association for Computational Linguistics, 2015.

[24] Y. Tang, Y. Huang, Z. Wu, H. Meng, M. Xu, and L. Cai. Question detection from acoustic features using recurrent neural network with gated recurrent unit. In *2016 IEEE International Conference on Acoustics, Speech and Signal Processing (ICASSP)*, pages 6125–6129, March 2016.

[25] Pete Warden. Research Blog: Launching the Speech Commands Dataset. Link: https://research.googleblog.com/2017/08/launching-speech-commands-dataset.html, August 2016.

[26] Wenpeng Yin, Katharina Kann, Mo Yu, and Hinrich Schütze. Comparative Study of CNN and RNN for Natural Language Processing. *arXiv:1702.01923 [cs]*, February 2017. arXiv: 1702.01923.